\documentclass[conference,letterpaper]{IEEEtran}

\usepackage[utf8]{inputenc}
\usepackage[T1]{fontenc}
\usepackage{url}
\usepackage{graphicx}

\title{A delay-based routing metric}
\author{
  \IEEEauthorblockN{Baptiste Jonglez}\IEEEauthorblockA{École Normale Supérieure de Lyon}
  \and
  \IEEEauthorblockN{Matthieu Boutier}
  \IEEEauthorblockA{Université Paris-Diderot}
  \and
  \IEEEauthorblockN{Juliusz Chroboczek}
  \IEEEauthorblockA{Université Paris-Diderot}
}

\begin{document}
\maketitle

\begin{abstract}
  In overlay networks, both local and long-distance links appear as
  a single hop to a routing protocol.  Traditional routing metrics
  (based on hop count or packet loss) fail to take the differences
  between such links into account.  In this paper, we study a metric
  based on packet delay that has been designed to improve routing in
  overlay networks.  We show a lightweight technique for measuring
  delay asynchronously, and show how to use the data it provides for
  constructing a routing metric.

  Using delay naively leads to persistent routing oscillations, so our
  routing protocol implements a number of features to bound the
  frequency of oscillations.  We show that our protocol causes no
  oscillations in real-world tests, and has oscillations with a period
  on the order of minutes in artificially constructed worst-case setups.
\end{abstract}

\section{Introduction}

An \emph{overlay network} is a network created on top of an existing
network.  In more technical terms, an overlay network is a network the
links of which are realised as flows (or connections) of the
underlying network.

Overlay networks are an old idea, and have many applications.  They
can be used as a transition technology, when the desired physical
network does not exist yet --- the transition to IPv6 was bootstrapped
by running IPv6 within the \emph{6bone}, an overlay over the existing
IPv4 Internet.  \emph{Virtual Private Networks} (VPN) are a technology
that allows a network node to appear connected at a place different
from what is implied by the physical network topology, typically in
order to work around topology-based security policies; \emph{onion
  routing} \cite{tor} generalises this idea to large public virtual
networks that are used to provide a modicum of anonymity to their
users.  Finally, by rerouting around failures faster than the
underlying network does, overlay networks are used to improve the
reliability of large-scale distributed systems in the presence of
partial network failures.  It is this last application that concerns
us here.

\subsection{Overlay networks for reliability}

BGP, the routing protocol used in the Internet core, is designed to
scale to very large networks.  This implies a number of trade-offs,
most notably relatively slow reconvergence after a network failure, on
the order of minutes.  Measurements indicate that at any one time
a few percent of the expected routes are not available \cite{detour}.
This implies that in a sufficiently large distributed system
implemented on the Internet, at any time at least some of the
participants will not be able to communicate.

There are multiple ways of dealing with this issue.  An interesting
approach is to design application algorithms that are able to deal
with temporary failures; for example, the SMTP protocol used for
electronic mail has a complex system of timeouts, retries and fallback
servers that allow it to deal with temporary failures.  A more recent
example is that of the Kademlia distributed hashtable algorithm (used
notably for locating peers in large-scale peer-to-peer file transfer
applications), which is highly redundant in order to deal with
arbitrary communication failures.

A more modular approach consists in delegating the reliability
requirements to a lower sub-layer.  In this approach, the application
blindly sends its data to the desired destination, and a lower
sub-layer uses an overlay network to route the data to the
destination, using a routing algorithm with fast rerouting properties
and with its own routing policies, possibly different from the
policies used by the underlying network.  This overlay network and
routing algorithm can be implemented within the application layer (as
an ad-hoc library), as in \emph{Resilient Overlay Networks}
\cite{ron}, which makes it possible to fine-tune the routing
heuristics in an application-specific manner (e.g.\ prefer lower
latency or higher reliability) without the need for cross-layer
interactions.  Alternatively, the overlay network can be implemented
at the network layer, using familiar packet-switching technology,
which reduces flexibility somewhat but allows using unmodified
applications over the overlay.

\subsection{Routing in a distributed cloud}

SlapOS is a framework for building distributed cloud applications.
SlapOS was initially implemented over native IPv6, which was found to
be too unreliable.  SlapOS was then modified to use a dense network
(but not a full mesh) of virtual links \cite{re6st}, and route over it
by using the off-the-shelf protocol Babel \cite{babel} with the
hop-count metric.

\begin{figure}[htb]
\centering
\includegraphics[width=0.35\textwidth]{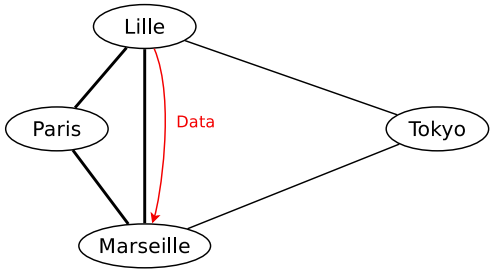}
\caption{A real-world topology}\label{fig:diamond-real}
\end{figure}

This solution worked fairly well as long as the cloud was mostly
local.  Unfortunately, as soon as distant nodes were added, Babel
started making routing choices that, while consistent with the
shortest-hop metric, were clearly sub-optimal.  Consider for example
the topology in Figure~\ref{fig:diamond-real}, which consists of four
nodes configured in an almost complete mesh.  As long as all the links
are operational, the shortest-hop metric yields optimal results ---
traffic local to Europe remains in Europe.  However, if the link
between Lille and Marseilles breaks, the shortest-hop metric does not
allow the routing protocol to distinguish between the local route
through Paris and the remote route through Tokyo, which is therefore
chosen in roughly one half of the cases.

The shortest-hop metric is not precise enough for the distributed
cloud.  In this paper, we describe our work on extending the Babel
routing protocol with a metric based on packet delay.

\subsection{A delay-based metric}

Our goal in this work is to extend the Babel routing protocol with the
simplest possible metric that does reliably distinguish between local
and non-local routes in the overlay network generated by SlapOS.  Our
metric is not meant to be the end-all of all metrics for overlay
networks; still, the requirements of the application dictate a number
of properties that it must have.

First, as one of the goals of the distributed cloud is to reduce
operational cost, the metric must not require any manual
configuration, which rules out manually configuring links as ``local''
or ``remote''.  We have chosen to base our network on the
\emph{round-trip time} (RTT), or two-way delay, which is easily
measured with off-the-shelf hardware with an accuracy sufficient to
distinguish between Paris and Tokyo.  (One-way delay might lead to
a more generally useful metric in the presence of asymmetric network
congestion, but it is more difficult to measure and is not required
for this particular application.)

Second, the algorithm must be easy to implement on cheap off-the-shelf
hardware, and, in particular, it must not rely on globally
synchronised clocks.  Since the links used in a distributed cloud are
of varying quality, it must consume a negligible amount of additional
network resources.  Additionally, since the hardware used in the
distributed cloud can be fairly loaded, it should be asynchronous,
i.e.\ not require real-time response to query packets.

Finally, since delay can be caused by network congestion, using delay
in a routing metric causes a feedback loop, which can cause persistent
oscillation.  We require that our algorithm provide reasonable
stability, with a bound on the period of oscillations of at least
a few minutes.

\subsection{Stability issues} \label{sec:stability}

Using delay as an input to the routing metric in congested networks
gives rise to a negative feedback loop: low RTT encourages traffic,
which in turn causes the RTT to increase.  In a discrete domain, such
a feedback loop can cause persistent oscillations.

\begin{figure}[htb]
\centering
\includegraphics[width=0.25\textwidth]{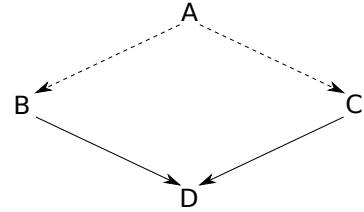}
\caption{Worst-case topology}
\label{fig:diamond-topology}
\end{figure}

Consider for example the topology in Figure~\ref{fig:diamond-topology},
where the links $A\cdot B$ and $A\cdot C$ are subject to congestion.
Suppose that there is a significant amount of traffic from $A$ to $D$.
The routing protocol initially chooses some route, say the route
through $B$; as the link $A\cdot B$ becomes congested, its RTT rises,
so the routing protocol reroutes through $C$.  The situation then
reverses: the link $A\cdot C$ becomes congested, the protocol reroutes
through $B$, etc.

In the general case, such oscillations are unavoidable in the presence
of congestion, but their frequency can be limited.  Our protocol
contains two mechanisms, saturation and hysteresis, that cooperate to
limit the frequency of oscillations; in Section~\ref{sec:worst-case},
we provide empirical data that shows that in the worst case the period
of the oscillations is on the order of minutes.

\section{Related work}

\subsection{Use of RTT in routing protocols}

In 1983, Mills described the use of RTT for routing in the
DCNet~\cite{mills83}, but didn't provide an evaluation of his
protocol; the asynchronous algorithm that we use to measure RTT
(Section~\ref{sec:async-rtt}) is inspired by Mills' algorithm, which
later became the basis for NTP~\cite{NTPv4}.  A few years later, the
``new'' routing protocol for the Arpanet~\cite{arpanet89} used
a metric based on RTT in order to mitigate the congestion of the
network; stability issues were considered, and solved by saturating
the metric, similarly to what we do.

Using a delay-based metric for routing has apparently been abandoned
since then: to the best of our knowledge, no modern network has been
using this method in recent years.  Our interpretation is that
congestion seldom occurs within the core of the network nowadays, and
has moved to the edge, where there is little opportunity for routing
optimisations: congestion occurs in the ``Customer Premises
Equipment'' (the ADSL modem) which cannot be routed around.

The proprietary routing protocols IGRP and EIGRP use a parameter
called ``delay'' for computing their metric.  However, this value is
statically configured by the operator rather than determined
empirically, and this feature is therefore out of scope for this
paper.

\subsection{Overlay networks}

Overlay networks are an old idea, and there is a wide range of
literature describing their various applications.  In this paper, we
are concerned with the use of overlay networks to increase
reliability, as described in Detour \cite{detour}.

The techniques most similar to ours are the ones used by
\emph{Resilient Overlay Networks} (RON) \cite{ron}, where the authors
build an overlay network to increase reliability and use a variety of
metrics, controlled by the application, to perform routing.  Unlike
our work, however, RON is layered above UDP and performs routing within
the application layer: this makes implementation simpler and makes it
easier to provide multiple routing metrics, but requires changing all
applications to link with the RON library and use its primitives for
communication.  In contrast to RON, our network-layer approach allows
the use of unmodified applications and is completely oblivious to the
transport-layer protocol being used.

\section{RTT-based routing}

In this section, we describe the issues related to integrating an
RTT-based metric in the Babel routing protocol.

\subsection{Measuring RTT asynchronously} \label{sec:async-rtt}

\begin{figure}[htb]
\begin{minipage}[b]{0.2\textwidth}
\centering
\includegraphics[scale=0.33]{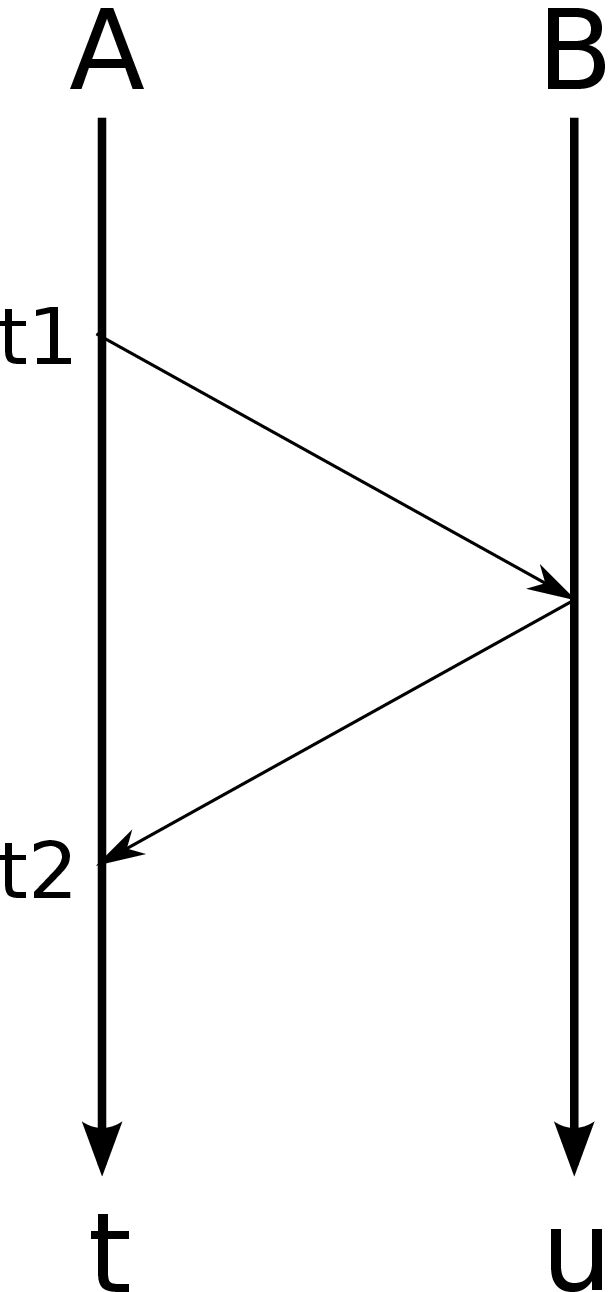}\\
(a)
\end{minipage}
\hfill
\begin{minipage}[b]{0.2\textwidth}
\centering
\includegraphics[scale=0.33]{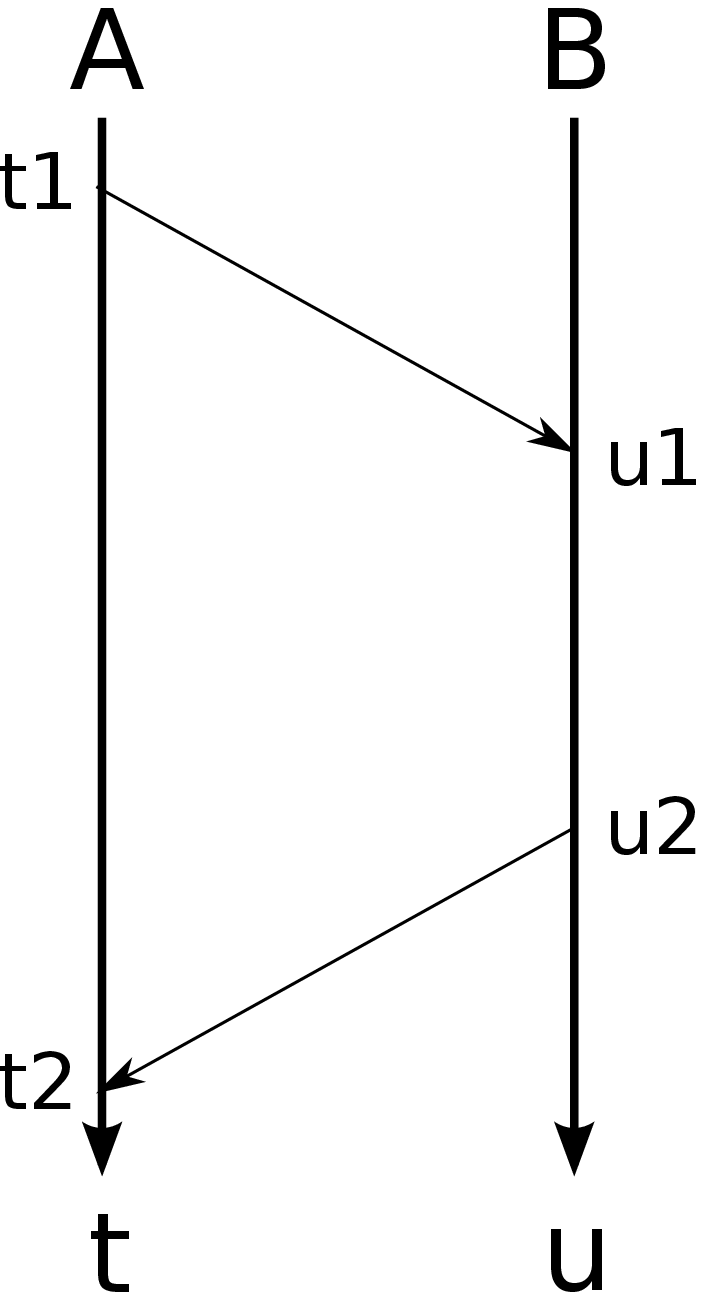}\\
(b)
\end{minipage}
\caption{RTT measurement}\label{fig:rtt}
\end{figure}

The simplest way to measure RTT between nodes A and
B (Figure~\ref{fig:rtt}(a)), as performed e.g.\ by the \emph{ping}
program, is to send a single ``echo request'' packet from A to B, and
have B immediately respond with an ``echo reply''.  This is a simple
and intuitive algorithm that does not require synchronised clocks;
unfortunately, it requires a synchronous reply from B, which is not
necessarily easy to integrate within an existing routing protocol.

Like most modern routing protocols, Babel has a fairly sophisticated
scheme for scheduling outgoing messages.  Roughly speaking, messages
are delayed by a random time (at most one \emph{Hello} interval) in
order to avoid global synchronisation \cite{jitter} and to make it
possible to aggregate multiple messages into a single packet.  Adding
synchronous messages to Babel would require a moderate amount of
changes to the protocol, increase the amount of network traffic it
generates, and might cause unexpected issues with node
synchronisation.

Fortunately, the problem of measuring RTT asynchronously has been
solved before, and was used by Mills in his \emph{HELLO} routing
protocol \cite{mills83} and in the NTP clock synchronisation protocol
\cite{NTPv4}.  In Mills' algorithm (Figure~\ref{fig:rtt}(b)), a node
A sends a packet $p_1$ with its local timestamp; $B$ saves $p_1$'s
reception timestamp $u_1$ according to its local clock.  At some later
time, a node B sends a packet $p_2$ with a copy $t_1$ of $p_1$'s
timestamp, its timestamp $u_1$, and the timestamp $u_2$ of $p_2$.
When node $A$ receives the packet $p_2$ at local time $t_2$, it
computes the difference
\[ (t_2 - t_1) - (u_2 - u_1) \]
which yields the RTT.  Note that each of the terms in this difference
use a single clock --- hence, no clock synchronisation is necessary.
Except for the first packet, all packets exchanged in Mills' algorithm
carry three timestamps: therefore, each node computes a new RTT sample
for each received packet, which is twice as efficient as the naive
\emph{ping} algorithm.

A further refinement is possible.  On a multi-access network,
a packet's timestamp is valid for all neighbours; it is only the
echoed timestamps which must be sent to a particular peer.  In Babel,
we attach a timestamp to each \emph{Hello} message, which is sent over
multicast to all neighbours.  The echoed timestamp is piggybacked to
\emph{IHU} (``I Heard You'') messages, used for reverse reachability
detection, which are conceptually unicast (but usually sent over
multicast).  In order to make it possible to perform Mills'
computation, we ensure that every \emph{IHU} is accompanied with
a \emph{Hello} in the same packet.  Therefore, the cost of
implementing Mills' algorithm is just a few octets per \emph{Hello}
and \emph{IHU} message, with no additional packets sent.

\subsection{Smoothing}

\begin{figure}[htb]
\centering
\includegraphics[width=0.45\textwidth]{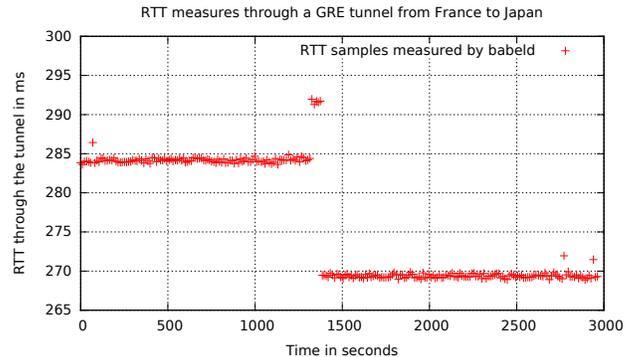}
\caption{RTT through a tunnel from Paris to
  Tokyo}\label{fig:rtt-paris-tokyo}
\end{figure}

The RTT samples obtained by the algorithm described above contain
a varying amount of \emph{jitter}, or short-term noise.
Figure~\ref{fig:rtt-paris-tokyo} shows the samples obtained over
a period of almost one hour over a GRE tunnel between Paris and Tokyo,
at a time when the RTT was particularly stable.  Before time 1300, the
samples are roughly constant, with a single outlier.  At time 1350,
something happens (rerouting?), there are a few outliers after which
time the samples are roughly constant again, with a small number of
outliers.

\begin{figure}[htb]
\centering
\includegraphics[width=0.45\textwidth]{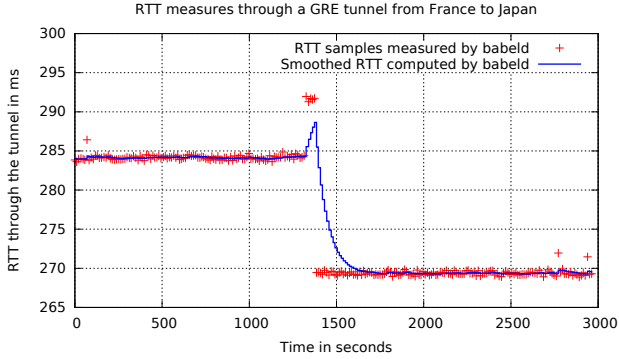}
\caption{Effect of smoothing on RTT}\label{fig:rtt-paris-tokyo-smoothed}
\end{figure}

Obviously, we are interested in the medium-term latency averages
(285\,ms before time 1500, 270\,ms after that), rather than in the
random jitter.  For that reason, we smooth the RTT data using an
exponential average analogous to the one used by TCP \cite{rfc793}.
More precisely, for every new RTT sample $\mathrm{RTT}_n$, our RTT
estimate $\mathrm{RTT}$ is updated as follows:
\[ \mathrm{RTT} := \alpha\cdot\mathrm{RTT} + (1 - \alpha)\cdot\mathrm{RTT}_n \]
The value $\alpha$ is currently set to 0.836 by default (which is
consistent with TCP's recommendation of 0.8 to 0.9).  The results of
this smoothing are shown in Figure~\ref{fig:rtt-paris-tokyo-smoothed}.

\begin{figure}[htb]
\centering
\includegraphics[width=0.45\textwidth]{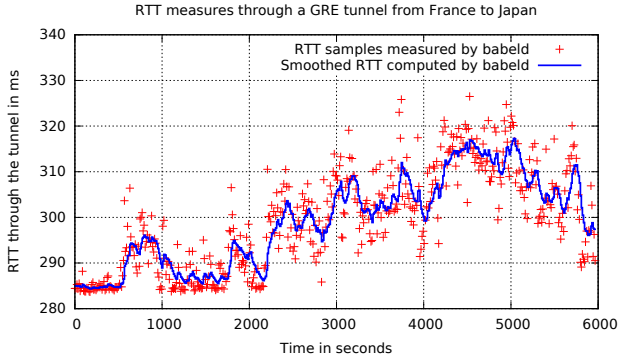}
\caption{Effect of smoothing on an unstable RTT}\label{fig:rtt-paris-tokyo-unstable}
\end{figure}

Figure~\ref{fig:rtt-paris-tokyo-unstable} shows the behaviour of the
same tunnel at a different time, when the RTT exhibited much larger
variation.  While the raw data is much more chaotic, the smoothing
algorithm is able to provide useful data.

\subsection{Accuracy and clock skew}

As noted above, Mills' algorithm does not require synchronised clocks.
However, its accuracy is limited by two factors.  First, packets must
be timestamped just before they are sent and just after they are
received: if sent packets are timestamped too early, or received
packets too late, the RTT will be overestimated.  Second, the two
clocks must progress at roughly the same rate: if one clock is
significantly faster than the other, RTTs will be overestimated on the
fast side and underestimated on the slow one.

Concerning the first issue, we have put some care into ensuring that
timestamps are generated in a timely manner.  Babel's packet formatter
formats \emph{Hello} messages with zero timestamps; the timestamps are
filled in just prior to transmission.  On the receiving side, however,
timestamps are only parsed after packet validation.  Our tests on
a local gigabit Ethernet indicate that we overestimate RTT by 0.4\,ms
as compared to the \emph{ping6} command, and introduce a moderate
amount of jitter, on the order of 0.1\,ms.  This is acceptable for the
intended application.

As to the second issue, a clock skew of $\delta$ introduces a maximum
error of $\delta\cdot\tau$, where $\tau$ is the maximum interval
between two IHU messages (12\,s by default).  Typical computer clocks
have clock skew on the order of 10\,ppm, which should yield an error
of at most 0.1\,ms.  Interestingly, our tests indicate that clock skew
increases dramatically when one peer enters a power-saving mode: in
that case, we have witnessed asymmetric errors of more than 1\,ms, an
order of magnitude more than the expected value.  Even these extreme
values, however, are within the accuracy required for the intended
application of our protocol.

\subsection{An RTT-based metric}

In the previous section, we described how to measure RTT precisely and
cheaply.  The RTT alone, however, does not directly constitute
a metric: we need to somehow map RTT values to an additive metric.

As far as the Babel routing protocol is concerned, a metric is just
a 16 bit integer.  While it would be possible to map RTT to a metric
proportionally (just multiplying it by some suitable constant), this
would favour too much low-RTT links, and prefer multiple low-RTT hops
to a single moderate-RTT hop.  What is more, it would yield
arbitrarily large metrics for large RTT links, which, as we shall see
in Section~\ref{sec:worst-case}, has a negative effect on stability.

\begin{figure}[htb]
\centering
\includegraphics[width=0.35\textwidth]{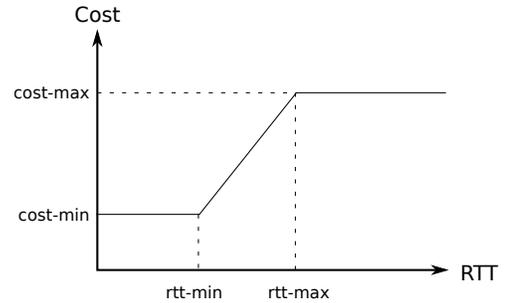}
\caption{Deriving cost from RTT}\label{fig:rtt-cost}
\end{figure}

Instead, we map RTT to metrics using the piecewise affine function
described in Figure~\ref{fig:rtt-cost}.  For RTTs below a value
\texttt{min-rtt} (10\,ms by default), a link is considered ``good'',
and its metric is the fixed value \texttt{min-cost}.  For RTTs above
\texttt{max-rtt} (120\,ms by default), the link is ``bad'', and its
cost is the fixed value \texttt{max-cost}.  For intermediate RTTs
between \texttt{min-rtt} and \texttt{max-rtt}, the resulting cost is
an affine function of the RTT.

This mapping has two essential properties.  First, all link metrics
are no smaller than \texttt{min-cost}, which guarantees that even very
low RTT links are not seen as ``free'' --- in a very low latency
network, our metric degenerates to the shortest-hop metric.  Second,
all high-RTT links are treated equally, which, as we shall see in
Section~\ref{sec:worst-case}, limits the frequency of route
oscillations in congested networks.

\subsection{Hysteresis}

In traditional routing protocols, metrics tend to vary
discontinuously, by discrete amounts.  Hence, a traditional routing
protocol can afford to switch routes as soon as a route's metric
becomes lower than that of the currently selected routes.  When
continuous metrics are used to measure real-world parameters, this is
no longer the case: the metrics of two routes could oscillate around
a similar value, leading to frequent route oscillation.  For that
reason, the Babel routing protocol applies a modest amount of
hysteresis to the metrics that it considers for route selection.  As
we shall see in Section~\ref{sec:worst-case}, this hysteresis is
essential to the stability of delay-based routing.

The algorithm is as follows.  For every route, Babel maintains two
metrics: the \emph{advertised metric} $M_a$, which is obtained from
neighbours and readvertised as is to other nodes, and the
\emph{smoothed metric} $M_s$.  The smoothed metric is initialised to
the advertised metric, and is periodically updated according to the
formula:
\[ M_s := \beta(\delta)\cdot M_s + (1 - \beta(\delta))\cdot M_a \]
where $\delta$ is the delay since the last update of $M_s$, and
$\beta(\delta)$ is a value chosen so that $M_s$ converges towards
$M_a$ exponentially with a time constant of 4\,s (in base 2).

Babel's route selection algorithm avoids routes with an infinite
advertised metric (retracted routes); when multiple routes to a given
destination have finite metrics, Babel will only switch routes if both
the advertised and smoothed metrics of the new route are better than
those of the currently selected one.  The effect is to permit fast
reconvergence when a route is lost, but to delay switching routes when
an unselected route's metric decreases below that of the currently
selected one.

The hysteresis algorithm may appear similar to the smoothing algorithm
described above, but there are good reasons why these are separate.
Babel is a modular protocol, and metric computation is separate from
route selection.  The smoothing algorithm is part of the metric
calculation, and is designed to extract a smooth signal from the noisy
RTT samples; it is specific to the RTT metric.  The hysteresis
algorithm, on the other hand, is part of the (metric-independent)
route selection procedure, and its only purpose is to improve
stability by delaying switching to a better route.

\section{Experimental evaluation}

In this section, we show some empirical data describing the behaviour
of our implementation of the algorithm described above.

\subsection{Real-world behaviour}

We have tested our implementation on a small overlay network deployed
over the Global Internet, consisting of four nodes, three of which are
in France and one in Japan.  The topology of the overlay network is
the one in Figure~\ref{fig:diamond-real}.  Each node is running Linux,
and the links are implemented using OpenVPN over UDP (without
cryptography).  All Babel instances are run with \verb|rtt-min| equal
to $10\,\textrm{ms}$, \verb|rtt-max| equal to $200\,\textrm{ms}$,
\verb|min-cost| equal to 96 and \verb|max-cost| equal to 246
Throughout the experiment, Lille is sending data to Marseilles.

\begin{figure}[htb]
\centering
\includegraphics[width=0.45\textwidth]{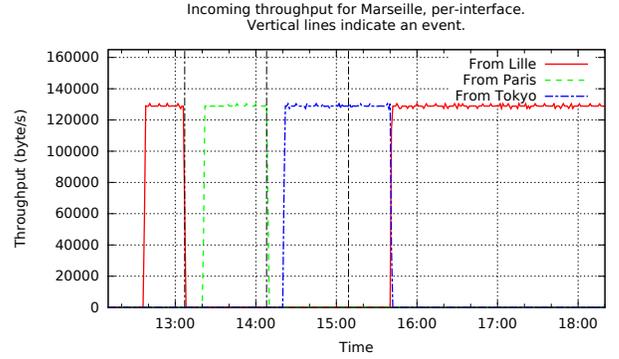}
\caption{Throughput in Marseilles}
\label{fig:nexedi-throughput}
\end{figure}

Figure~\ref{fig:nexedi-throughput} shows the incoming throughput in
Marseilles over each of the local interfaces.  Initially, all links
are up, so the data arrives directly from Lille.  Around minute 13,
the direct link between Lille and Marseilles is shut down; after a few
dozen seconds, the failure is detected, and the data is rerouted
through Paris.  Around minute 14, the Paris link is shut down, and the
data is rerouted through Tokyo.  Finally, after minute 1d, the links
are reestablished; when this is detected, the data is rerouted through
the direct low-latency link.

\begin{figure}[htb]
\centering
\includegraphics[width=0.45\textwidth]{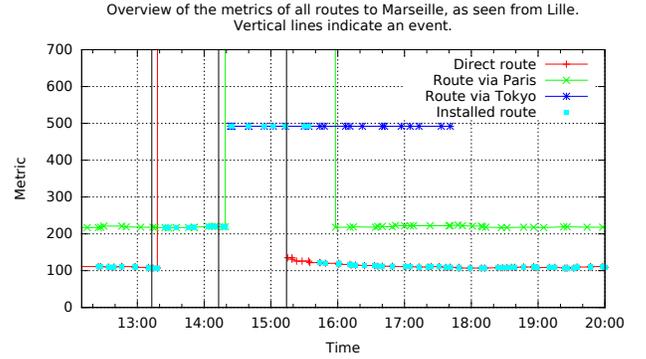}
\caption{Metrics in Lille}
\label{fig:nexedi-metrics}
\end{figure}

Figure~\ref{fig:nexedi-metrics} shows the metrics of the different
routes during the experiment.  It shows that the links remain
uncongested: all of the metrics remain roughly constant throughout the
experiment.

\subsection{Worst-case simulation} \label{sec:worst-case}

The previous experiment uses links of different natural latencies that
remain uncongested throughout the experiment.  We believe that this is
representative of real-world conditions in overlay networks; however,
since the traffic that we generate does not significantly impact the
latencies of the links, the feedback loop described in
Section~\ref{sec:stability} does not occur, and there are no stability
issues.

In order to test our algorithm's stability properties in a worst case
situation, we have simulated a network consisting of two exactly
identical parallel routes that are subject to congestion.  The
topology is that of Figure~\ref{fig:diamond-topology}; the links $A\cdot
B$ and $A\cdot C$ have their throughput artificially limited, and are
therefore subject to congestion, while the links $B\cdot D$ and
$C\cdot D$ are uncongested.

\begin{figure}[htb]
\centering
\includegraphics[width=0.45\textwidth]{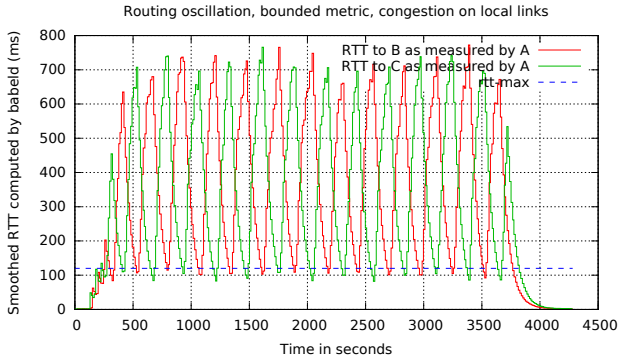}
\caption{Worst-case RTT oscillation}
\label{fig:diamond-stability-rtt}
\end{figure}

As expected, routing in this somewhat pathological topology is subject
to oscillations.  Figure~\ref{fig:diamond-stability-rtt} shows the RTTs
of the two congested links.  The routing protocol chooses one of the
two routes, the RTT of which subsequently increases; after a few
minutes, the protocol reacts to the increase of the RTT and switches
to the other route; the situation then repeats, \emph{ad nauseam}.
However, the frequency of the oscillations remains bounded, with
a time constant of roughly 5 minutes.

Two mechanims collaborate to limit the frequency of oscillations.  The
saturation of the cost function ensures that both congested links
spend part of their time in the saturated state.  Hysteresis ensures
that Babel doesn't switch routes as long as both metrics are saturated.

\begin{figure}[htb]
\centering
\includegraphics[width=0.45\textwidth]{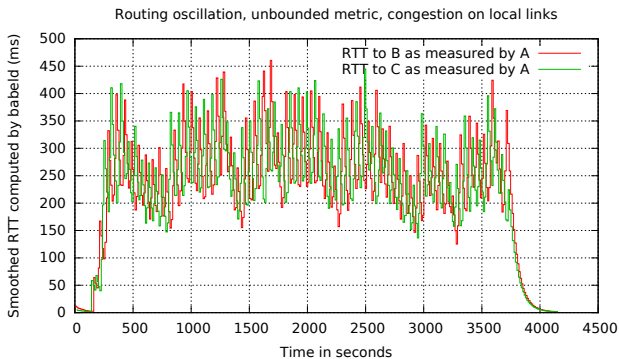}
\caption{Worst-case RTT oscillation without saturation}
\label{fig:diamond-stability-rtt-unbounded}
\end{figure}

Figure~\ref{fig:diamond-stability-rtt-unbounded} shows an experiment
performed in the same topology, but with an unbounded cost function
(both \verb|rtt-max| and \verb|cost-max| set to very high values,
chosen so that the slope of the curve remains the same as in the
previous experiment).  The oscillations are now much faster (less than
a minute), which shows the importance of a bounded cost function.

In this experiment, the congested links are the ones close to the
sender, which ensures fast reaction to changing conditions.  We have
repeated the experiment with the links $B\cdot D$ and $C\cdot D$ being
the ones subject to congestion; as expected, the behaviour is similar,
but with slightly slower oscillation.

\section{Conclusions and further work}

In this paper, we have described a working implementation of
a delay-based routing metric that is currently being deployed in
production.  We have shown an algorithm that measures RTT while having
a negligible impact on the amount of routing protocol traffic, and
have shown how to mitigate the stability issues intrinsically
connected to using delay that are good enough to limit instability in
the most hostile examples that we could construct.

While the functionality of our protocol is sufficient for the overlay
networks that we consider, there is a number of related issues that
still remain open.

\paragraph{One-way delay}

The metric described in this paper is based on the round-trip time, or
two-way delay.  The congestion control community have repeatedly shown
that one-way delay behaves better than two-way delay, at least as far
as congestion control algorithms are concerned \cite{tcp-lp,ledbat},
at the cost of much more complex algorithms.  It would certainly be
interesting to find out whether there are any real-world cases where
one-way delay performs significantly better than RTT as a basis for
a routing metric.

\paragraph{Arbitrary choices and theoretical study of stability}

There are a number of arbitrary choices in our algorithm: the
constants used for smoothing and filtering, the amount of hysteresis
applied, and, above all, the function used for mapping an RTT value to
a metric.  While we have empirically checked that these particular
choices work well, at least for the particular application under
consideration, there are almost certainly other choices that would
work just as well and perhaps better.

More generally, we lack an in-depth theoretical understanding of the
performance of our algorithm, in particular of its stability.  As
there exist a number of techniques for the theoretical study of the
stability of distributed systems, this would seem to be feasible.

\paragraph{Other applications}

After we initially published the code of our implementation, one
researcher has expressed interest in studying its suitability for
networks other than overlays.  There is some support to the feeling
that the metrics currently used in wireless mesh networks (such as ETX
\cite{etx} or physical-layer based metrics) are not satisfactory,
because they are a poor predictor of network performance, because they
are too slow to react to changing conditions, or because they are too
difficult to implement.  We hold some hope that, at least for some MAC
layers, an accurate measurement of delay might be a good indicator of
lower-layer congestion, and therefore could serve as one component of
a metric for wireless mesh networks.

\section*{Acknowledgment}

We are grateful to Julien Muchembled of Nexedi for providing access to
the cloud nodes used in our experiments.

\end{document}